\documentclass[letterpaper, 10pt, conference]{ieeeconf}

\IEEEoverridecommandlockouts 

\usepackage{times}
\usepackage{multicol}
\usepackage[bookmarks=true]{hyperref}

\usepackage{graphicx}
\usepackage{eurosym}
\usepackage{array}
\usepackage{subfigure}
\usepackage{amssymb,amsmath,amsfonts}
\usepackage{booktabs}
\usepackage{multirow}
\usepackage[table]{xcolor}
\usepackage{arydshln}
\usepackage{bbm}
\usepackage{cite}

\usepackage{algorithm}
\usepackage[noend]{algpseudocode}

\usepackage{float}
\usepackage{psfrag}
\usepackage{hyperref}
\hypersetup{
    colorlinks,
    citecolor=blue,
    filecolor=blue,
    linkcolor=blue,
    urlcolor=blue,
    pdfstartview={XYZ null null 1.00}
}

\definecolor{gray1}{gray}{0.7}
\definecolor{gray2}{gray}{0.9}

\newcommand{\decBayes}{\textit{Bayes-Swarm}}
\newcommand{\alphaCoef}{exploitation weight}
\newcommand{\bx}{\mathbf{x}}
\usepackage{dblfloatfix}    

\begin{document}

\title{\LARGE \bf
Informative Path Planning with Local Penalization for Decentralized and Asynchronous Swarm Robotic Search
}

\author{Payam Ghassemi$^{1}$ and Souma Chowdhury$^{2}$
\thanks{$^{1}$Ph.D. Student, Dept. of Mechanical and Aerospace Engineering, University at Buffalo, Buffalo, NY 14260, USA.
        {\tt\small payamgha@buffalo.edu}}%
\thanks{$^{2}$Assistant Professor, Dept. of Mechanical and Aerospace Engineering, University at Buffalo, Buffalo, NY 14260, USA. \textit{Corresponding Author}. 
        {\tt\small soumacho@buffalo.edu}}%
}

\maketitle

\begin{abstract}
Decentralized swarm robotic solutions to searching for targets that emit a spatially varying signal promise task parallelism, time efficiency, and fault tolerance. It is, however, challenging for swarm algorithms to offer scalability and efficiency, while preserving mathematical insights into the exhibited behavior. A new decentralized search method (called {\decBayes}), founded on batch Bayesian Optimization (BO) principles, is presented here to address these challenges. Unlike swarm heuristics approaches, {\decBayes} decouples the knowledge generation and task planning process, thus preserving insights into the emergent behavior. Key contributions lie in: 1) modeling knowledge extraction over trajectories, unlike in BO; 2) time-adaptively balancing exploration/exploitation and using an efficient local penalization approach to account for potential interactions among different robots' planned samples; and 3) presenting an asynchronous implementation of the algorithm. This algorithm is tested on case studies with bimodal and highly multimodal signal distributions. Up to 76 times better efficiency is demonstrated compared to an exhaustive search baseline. The benefits of exploitation/exploration balancing, asynchronous planning, and local penalization, and scalability with swarm size, are also demonstrated.
\end{abstract}

\vspace{0.1cm}
\begin{keywords}
Swarm Robotic Search, Informative Path Planning, Bayesian Search, Gaussian Process, Asynchronous.
\end{keywords}

\IEEEpeerreviewmaketitle


\section{INTRODUCTION} \label{sec:intro}
Swarm robotic search is concerned with searching for or localizing targets in unknown environments with a large number of collaborative robots. There exists a class of search problems in which the goal is to find the source or target with maximum strength (often in the presence of weaker sources), and where each source emits a spatially varying signal. Potential applications include source localization of gas leakage~\cite{baetz2009mobile}, nuclear meltdown tracking~\cite{nagatani2013emergency}, chemical plume tracing~\cite{li2006moth}, and magnetic field and radio
source localization~\cite{viseras2016decentralized,song2012simultaneous}.
In such applications, decentralized swarm robotic systems have been touted to provide mission efficiency, fault tolerance, and scalable coverage advantages~\cite{desilva2012development,ghassemi2018decentralized,ghassemi2019decentralized}, compared to sophisticated standalone systems. Decentralized search subject to a signal with unknown spatial distribution usually requires both task inference and planning, which must be undertaken in a manner that maximizes search efficiency and mitigates inter-robot conflicts. This in turn demands decision algorithms that are computationally light-weight (i.e., amenable to onboard execution) ~\cite{pugh2007inspiring}, preferably explainable~\cite{gunning2017explainable}, and scalable~\cite{tan2013research} -- it is particularly challenging to meet these characteristics simultaneously. 

In this paper, we perceive the swarm robotic{\footnote{Note that, we take a broader perspective of self-organizing swarm systems, one that is \textbf{not} limited to superficially-observable pattern-formations.}} search process to consist of creating/updating a model of the signal environment and deciding future waypoints thereof. Specifically, we design, implement and test a novel decentralized algorithm founded on a batch Bayesian search formalism. This algorithm tackles the balance between exploration and exploitation over trajectories (as opposed to over points, which is typical in non-embodied search), while allowing asynchronous decision-making. 
The remainder of this section briefly surveys the literature on swarm search algorithms, and converges on the contributions of this paper.
%
\subsection{Swarm Robotic Search}
In time-sensitive search applications under complex signal distributions, 
a team of robots can broaden the scope of operational capabilities through distributed remote sensing, scalability and parallelism (in terms of task execution and information gathering)~\cite{odonkor2019distributed}. The \textit{multi-robot search} paradigm~\cite{tan2013research} uses concepts such as cooperative control~\cite{sinha2017cooperative}, model-driven strategies,~\cite{wiedemann2017bayesian}, Bayesian filter by incorporating mutual information~\cite{charrow2014cooperative}, strategies based on local cues~\cite{hajieghrary2016multi}, and uncertainty reduction methods \cite{sujit2009negotiation}. Scaling these methods from the multi-robotic ($<$10 agents ~\cite{tan2013research}) to the swarm-robotic level (10 to 100 agents) often becomes challenging in terms of online computational tractability.


A different class of approaches that is dedicated to guiding the search behavior for larger teams is that based on nature-inspired \textit{swarm intelligence} (SI) principles ~\cite{kennedy2010particle,krishnanand2009glowworm,senanayake2016search}. SI-based heuristics have been used to design algorithms both for search in non-embodied $n$-dimensional space (e.g., particle swarm optimization) and for swarm robotic search \cite{kennedy2006swarm, bonabeau1999swarm}. Majority of the latter methods are targeted at localizing a single source \cite{pugh2007inspiring,jatmiko2006pso}, with their effectiveness relying on the usage of adaptive parameters (e.g., changing inertia weight)~\cite{jatmiko2006pso}. 
The localization of the maximum strength source in the presence of other weaker sources (i.e., under a multi-modal spatial distribution), without making limiting assumptions (e.g., distributed starting points \cite{krishnanand2006glowworm}), has received much less attention among SI-based approaches. 

\textit{Translating optimization processes}: Similar in principle to some SI approaches, here we aim to translate an optimization strategy~\cite{ghassemi2019adaptive} to perform search in the physical 2D environment. In doing so, it is important to appreciate two critical differences between these processes: \textbf{1)} \textit{Movement cost:} unlike optimization, in swarm robotic search, moving from one point to another may require a different energy/time cost depending upon the environment (distance, barriers, etc.) separating the current and next waypoints. \textbf{2)} \textit{Sampling over paths:} robots usually gather multiple samples (signal measurements) over the path from one waypoint to the next (as sampling frequency $>>$ waypoint frequency), unlike in optimization where we sample only at their next planned point. 
This ``sampling over paths" characteristic has received minimal attention in existing SI-based approaches. 

Moreover, with SI-based methods, the resulting \textit{emergent} behavior, although often competitive, raises questions of dependability (due to the use of heuristics) and mathematical explainability~\cite{kolling2016human}). 
The search problem can be thought of as comprising two main steps: task inference (identifying/updating the signal spatial model) and task selection (waypoint planning). In SI methods, the two steps are not separable, and a spatial model is not explicit. In our proposed approach, the processes are inherently decoupled -- robots exploit Gaussian Processes to model the signal distribution knowledge (task inference) and solves a 2D optimization over a special acquisition function to decide waypoints (task selection). Such an approach is expected to provide explainability, while preserving computational tractability. 
\vspace{-0.1cm}

\subsection{Contributions of this Paper}
%
The primary contributions of this paper, comprising what we call the {\decBayes} algorithm, can be summarized as the following. {\textbf{1)} We extend Gaussian process modeling to update over trajectories and consider robot's motion constraints when using the GP to identify new samples. \textbf{2)} We formulate a novel batch-BO acquisition function, which not only seeks to balance exploration and exploitation, but also manages the interactions between samples in a batch, i.e., different robots' planned future paths, in a computationally efficient manner. \textbf{3)} We develop and test a simulated parallel implementation of {\decBayes} for asynchronous search planning over complex multi-modal signal distributions.} The performance of {\decBayes}, and its variations (synchronized and purely explorative implementations), is analyzed over signal environments of different complexity, and compared with that of an exhaustive search baseline.

The remaining portion of the paper is organized as follows: Our proposed decentralized algorithm (\decBayes) is described next in Section~\ref{sec:decBayes}. Numerical experiments and results, investigating the performance of {\decBayes} for different case studies, are then presented in Sections \ref{sec:expts} and \ref{sec:results}. The paper ends with concluding remarks. A summary background of GP modeling is provided as Appendix \ref{sec:GP} thereafter.

\section{BAYES-SWARM ALGORITHM}\label{sec:decBayes}
\vspace{-0.1cm}
\subsection{Bayes-Swarm: Overview}
\begin{figure}[!t]
\centering
\includegraphics[width=0.42\textwidth]{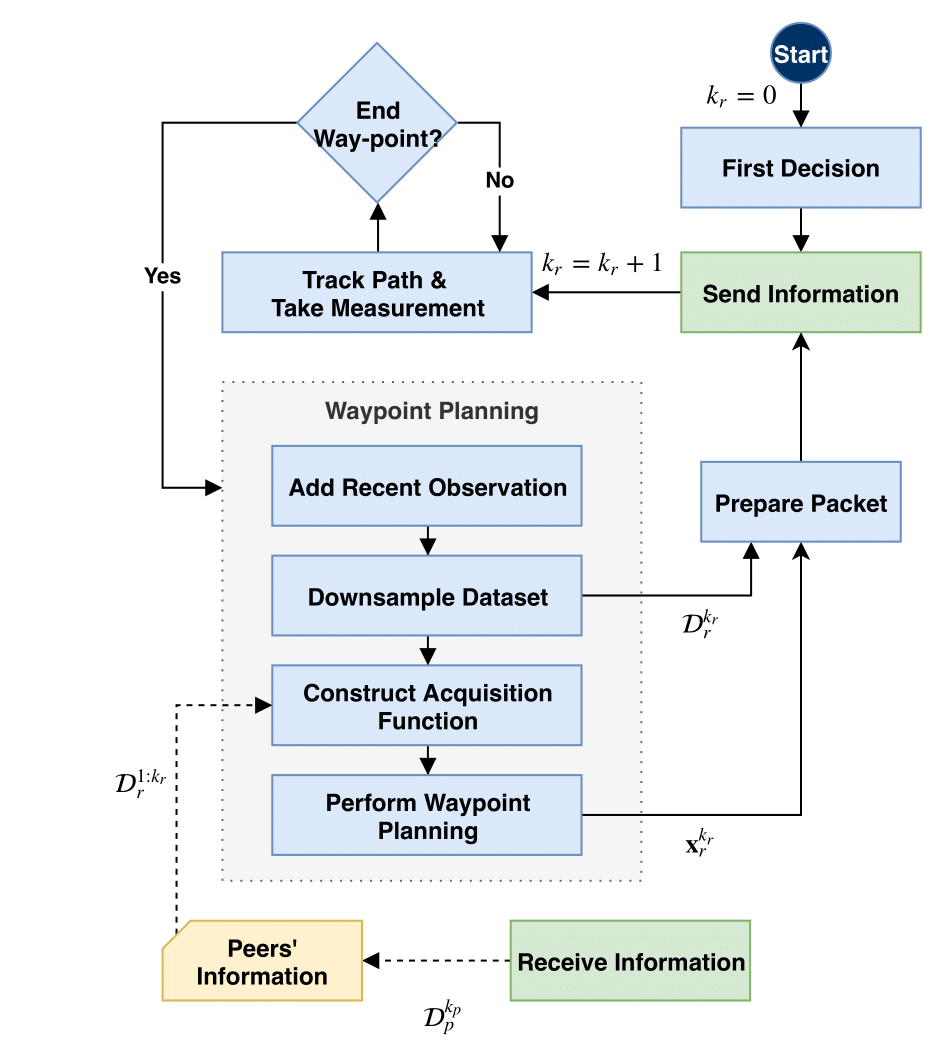}
\vspace{-0.3cm}
\caption{\decBayes: sequence of processes for each robot in the swarm}
\label{fig:decBayesflowchart}
\vspace{-0.6cm}
\end{figure}

Figure ~\ref{fig:decBayesflowchart} illustrates the sequence of processes (motion, sensing, planning, communication, etc.), and associated flow of information, encapsulating the behavior of each swarm robot. The pseudocode of our proposed \decBayes~algorithm is given in Alg.~\ref{alg:decBayes}. Each robot executes the \decBayes~algorithm after reaching a waypoint, to update its knowledge and identify the next waypoint. This is accomplished by updating the GP model of the signal environment, and using it to maximize a special acquisition function. Importantly, these planning instances need not be synchronized across robots. The following assumptions are made in implementing {\decBayes}: i) all robots are equipped with precise localization; and ii) each robot can communicate their knowledge and decisions to all peers after reaching a waypoint (while full observability is assumed, the provision of asynchronous planning reduces the communication dependency, compared to synchronized algorithms~\cite{klavins2004communication,cap2013asynchronous}.

Assuming we have $m$ robots, let's define the key parameters in {\decBayes}: $\mathcal{D}_r^{k_r}=[\mathbf{X}_r^{k_r}, \mathbf{y}_r^{k_r}]$: the observation locations ($\mathbf{X}_r^{k_r}$) and signal measurements ($\mathbf{y}_r^{k_r}$) of robot-$r$ over its path connecting waypoints $k_r$ and $k_{r-1}$; $\mathcal{D}^{1:k_r}=\bigcup_{r=1}^{m}\bigcup_{i=1}^{k_r}\mathcal{D}_r^i$: the cumulative information of robot-$r$ up to its arrival at the $k_r$-{th} waypoint, including all self-recorded and peer-reported observations; $\hat{\mathbf{x}}_{p}^{k_r}$: the next planned waypoint of robot-$p$, known to robot-$r$ at the time when it's at its $k_r$-th waypoint; $\mathbf{\hat{X}}_{-r}^{k_r} = \bigcup_{p=1 \wedge p\neq r} \hat{\mathbf{x}}_{p}^{k_r}$: reported next waypoints of robot-$r$'s peers by that time.
\begin{algorithm}[!b]
\caption{\decBayes~Algorithm}\label{alg:decBayes}
\small
\textbf{Input:} $GP_r, \mathbf{x}_r, X^{k_r}_{-r}$ -- current waypoint and recent observations of robot-$r$ ($\mathbf{x}$), and planned waypoints of peers ($X^{k_r}_{-r}$).\\
\textbf{Output:} $x_r^{k_r+1}$ - the next waypoint of robot-$r$.
\begin{algorithmic}[1]
\Procedure{takeDecision}{$r, k_r, m, \Delta\theta$}
\If {$k_r = 0$}
\State {$\mathbf{x}_r^{k_r}\gets $ \textsc{takeFirstDecision}($r, k_r, m, \Delta\theta$)}
\Else
\If {Size of $\mathcal{D}_r^{k_r} > N_\text{max}$}\Comment{$N_\text{max} = 1,000$}
\State {Down-sample $\mathcal{D}_r^{k_r}$ to $N_\text{max}$ $n$ observations}
\EndIf
\State {$\mathbf{\hat{x}}_{0} \gets $ Location with the highest observed value}
\State {$\mathbf{x}_r^{k_r} \gets $ Location maximizing the acquisition func., Eq.\eqref{eq:mainObjectiveFunc}, by a gradient-based solver w/ initial $\mathbf{\hat{x}}_{0}$}
\EndIf
\State {$k_r \gets k_r + 1$}
\State {\textbf{return} {$\mathbf{x}_r^{k_r}, k_r$}}
\EndProcedure
\Procedure{takeFirstDecision}{$r, m, \Delta\theta, V_r, T$}
\State {$d \gets V T$}
\If {$\Delta\theta = 360$} \Comment{$\Delta\theta$: Initial feasible dir. range}
\State {$\theta \gets r\Delta\theta/m$}
\Else
\State {$\theta \gets r\Delta\theta/(m+1)$}
\EndIf
\State {$\mathbf{x}_r^1 \gets [d \cos\theta, d \sin\theta]$}
\State {\textbf{return} {$\mathbf{x}_r^1$},}
\EndProcedure
\end{algorithmic}
\end{algorithm}
%

\subsection{Acquisition Function}
Each robot-$r$ takes an action (plans/travels-to next waypoint, $x_r^{k_r+1}$) that maximizes an acquisition function. Since the swarm's objective is to collectively explore a search area to find the strongest signal source among multiple sources in the least amount of time, the acquisition function must balance exploration and exploitation. To this end, we propose the following acquisition function formulation:
\begin{equation}
\vspace{-0.4cm}
\label{eq:mainObjectiveFunc}
\mathbf{x}^{k_r+1}_r = \text{arg}\max_{\mathbf{x}\in \mathcal{X}^{k_r}} ( \alpha \cdot \Omega_r + (1 - \alpha) \beta \Sigma_r ) \; \Gamma_r
\end{equation}
\vspace{-0.1cm} s.t. \vspace{-0.1cm}
\begin{equation}
\label{eq:constraint}
0\leq l_s^{k_r}=\|\mathbf{x} - \mathbf{x}_r^{k_r}\|\leq V T
\end{equation}
Here, the first term, $\Omega_r = \Omega_r(\mathbf{x}, \mathcal{D}^{1:k_r})$, leads robot-$r$ towards the location of the maximum signal strength expectation (promotes exploitation); and the second term, $\Sigma_r = \Sigma_r(\mathbf{x}, \mathcal{D}^{1:k_r},\mathbf{\hat{X}}_{-r}^{k_r})$, minimizes the knowledge uncertainty of robot-$r$ w.r.t. the signal's spatial distribution (promotes exploration). The multiplicative factor, $\Gamma(\mathbf{x}, \mathbf{\bar{X}}_{-r}^{k_r})$, penalizes the interactions among the samples planned to be collected by robot-$r$ and its peers. How these terms are formulated differently from standard acquisition functions used in BO (to enable the unique characteristics of embodied search), are described further in the following sub-sections.

The coefficient $\alpha\in\left[0,1\right]$ in Eq. \eqref{eq:mainObjectiveFunc} is the \alphaCoef, i.e., $\alpha=1$ would be purely exploitative. Here, we design $\alpha$ to be adaptive in a way (given by Eq. \eqref{eq:alpha-adapt}) such that the swarm behavior is strongly explorative at the start and becomes increasingly exploitative over waypoint iterations, e.g., it changes from $\approx$0.5 to $\approx$0.97 between 33\% and 70\% of the expected mission time:
\vspace{-0.2cm}
\begin{equation}\label{eq:alpha-adapt}
    \alpha = 1/\big(1+\exp(-10(\frac{t}{T_\text{max}}-\frac{1}{3}))\big)
\vspace{-0.2cm}
\end{equation}
The term $\beta$ in Eq.~\eqref{eq:mainObjectiveFunc} is a prescribed scaling parameter that is used to align the orders of magnitude of the exploitative and explorative terms. For our case studies, we set $\beta=50$. Equation \eqref{eq:constraint} constrains the length of robot-$r$'s planned path, $l_s^{k_r}$), based on a set time-horizon ($T$) for reaching the next waypoint, and the maximum velocity ($V$) of robots. In this paper, the time-horizon is set such that $6\times$distance between any consecutive waypoints does not exceed the arena length.  

\subsection{Source Seeking (Exploitative) Term}
The robots model the signal's spatial distribution using a GP with squared exponential kernel (further description of this GP modeling is given in Appendix \ref{sec:GP}). The GP model is updated based on the robot's own measurements and those communicated by its peers (each over their respective most recent paths), thereby providing the following mean function:
\vspace{-0.2cm}
\begin{equation}
    \mu_r(\mathbf{x}) = \mu(\mathbf{x},\mathbf{X},\mathbf{y})
\vspace{-0.1cm}
\end{equation}
where $\mathbf{X}=\bigcup_{r=1}^{m}\bigcup_{i=1}^{k_r}{\mathbf{X}_r^i}$ and $\mathbf{y} = \bigcup_{r=1}^{m}\bigcup_{i=1}^{k_r}y_r^i$.
Due to motion constraints (Eq.~\eqref{eq:constraint}, a robot may not be able to reach the location, $\mathbf{\bar{x}}^*$, with the maximum expected signal strength (estimated using their GP model), within the time horizon. Therefore, the exploitative term is re-defined to get closest to $\mathbf{\bar{x}}^*$, as given by:
\begin{align}
\label{eq:sourceSeeking}
\Omega_r(\mathbf{x},\mathcal{D}) &= \frac{1}{1+(\mathbf{x} - \mathbf{\bar{x}}^*)^T(\mathbf{x} - \mathbf{\bar{x}}^*)}
\end{align}
where $\mathbf{\bar{x}}_r^* = \text{arg}\max_{\mathbf{\bar{x}}}\mu_r(\mathbf{\bar{x}})$.
\subsection{Knowledge-Uncertainty Reducing (Explorative) Term}
Unlike in optimization, in robotic search, sampling is performed over the path of each agent. This concept is known as informative path planning, where robots decide their path such that the best possible information is extracted. The (explorative) second term in Eq. \eqref{eq:mainObjectiveFunc} models the reduction in uncertainty in the robots' belief (knowledge), thus facilitating informative path planning. To this end, the path of the robot is written in a parametric form as given below:
\begin{align}
s(u) &= u \mathbf{x} + (1-u)\mathbf{x}_r^{k_r};\; u\in[0,1]
\end{align}
where $\mathbf{x}_r^{k_r}$ is the current location of robot-$r$. In computing the self-reducible uncertainty in the belief of robot-$r$, we account for the locations of both the past observations made by the robot and its peers and the future observations of robot-$r$'s peers to be made over the paths to their planned waypoints ($\mathbf{\bar{X}}_{-r}^{k_r}$) -- both of these only consider what's currently known to robot-$r$ via communication from its peers. The knowledge-gain can thus be expressed as:
\vspace{-0.22cm}
\begin{equation}
\label{eq:knowledgeGain}
\Sigma_r(\mathbf{x}, \mathcal{D}^{1:k_r},\mathbf{\bar{X}}_{-r}^{k_r}) = \int_{s(\mathbf{x})} \sigma_r(s(u)) du
\vspace{-0.22cm} 
\end{equation}
%
where $\sigma_r(\mathbf{x}) = \sigma(\mathbf{x},\mathbf{X}_r^e)$ and $\mathbf{X}_r^e = \mathbf{X}\cup\mathbf{\bar{X}}_{-r}^{k_r}$. 
For further details on computing the mean (Eq. \eqref{eq:sourceSeeking}) and the variance (Eq. \eqref{eq:knowledgeGain}) of the GP, refer to Appendix \ref{sec:GP}. 

\subsection{Local Penalizing Term}
For a batch-BO implementation, it is necessary to account for (and in our case mitigate) the interaction between the batch of future samples. In swarm robotic search, this provides an added benefit of mitigating the overlap in planned knowledge gain by robots in the swarm -- thereby promoting a more efficient search process. Modeling it explicitly via predictive distribution carries a significant computational overhead of $O(n^3)$ \cite{gonzalez2016batch}. Simultaneous optimization of future candidate samples (in the batch) \cite{azimi2010batch} is also not applicable here since each robot must plan its future waypoint in a decentralized manner. Recently, Gonzalez et al.~\cite{gonzalez2016batch} reported a computationally tractable approximation to model the interactions, using a local penalization term. We adopt and extend this idea in our work through the local penalty factor.  

This penalty factor, $\Gamma(\mathbf{x}, \mathbf{\hat{X}}_{-i}^{p})$ enables local exclusion zones based on the Lipschitz properties of the signal spatial function ($f(x)$), and thus tends to smoothly reduce the acquisition function in the neighborhood of the existing batch samples (the known planned waypoints of robot-$r$'s peers ($\mathbf{\hat{X}}_{-r}^{k_r}$), with the signal observations at those points being not yet reported). To compute this penalty, we define a ball $\mathcal{B}_r$ with radius $\rho$ around each peers' planned waypoints:
\vspace{-0.1cm}
\begin{equation}
   \mathcal{B}_r(\mathbf{x},\mathbf{\hat{x}}_p^{k_p})
    = \{\mathbf{x}\in\mathcal{X}:\;\|\mathbf{\hat{x}}_p^{k_p} - \mathbf{x}\| \leq \rho \}; ~
    \mathbf{\hat{x}}_p^{k_p}\in \mathbf{\hat{X}}_{-r}^{k_r}
\end{equation}
The local penalty associated with a point $\mathbf{x}$ is defined as the probability that $\mathbf{x}$ does not belong to the ball $\mathcal{B}_r$:
\vspace{-0.1cm}
\begin{equation}
\label{eq:localPenalizer}
    \gamma(\mathbf{x},\mathbf{\hat{x}}_p^{k_p}) = 1 - P(\mathbf{x}\in \mathcal{B}_r)
\end{equation}
We assume that the distribution of the ball radius $\rho$ is Gaussian with mean $(M-\mu_r(\mathbf{\hat{x}}_p^{k_p}))/ L$ and variance $\sigma_r^2(\mathbf{\hat{x}}_p^{k_p})/L^2$. Here, $M = \max_{\mathbf{x}} f(\mathbf{x})$ is the maximum strength of the source signal and $L$ is a valid Lipschitz constant ($\|f(\mathbf{x}_1)-f(\mathbf{x}_2)\|
\leq L\|\mathbf{x}_1-\mathbf{x}_2\|$). Both $M$ and $L$ can be in general readily set based on the knowledge of the application (we know the expected maximum strength of the source signal and the size of arena). By having these assumptions, we can derive the following expressions for the local penalty:
\begin{equation}
    \begin{aligned}
    \gamma(\mathbf{x},\mathbf{\hat{x}}_p^{k_p}) &= 1 - P(\|\mathbf{\hat{x}}_p^{k_p} - \mathbf{x}\|\leq \rho)\\
    &=P(\mathcal{N}(0,1)\leq\frac{L\|\mathbf{x}-\mathbf{\hat{x}}_p^{k_p}\|-M+\mu_r(\mathbf{\hat{x}}_p^{k_p})}{\sigma_r(\mathbf{\hat{x}}_p^{k_p})})\\
    &=\frac{1}{2}\text{erfc}\left(-\frac{L\|\mathbf{x}-\mathbf{\hat{x}}_p^{k_p}\|-M+\mu_r(\mathbf{\hat{x}}_p^{k_p})}{\sqrt{2\sigma_r^2(\mathbf{\hat{x}}_p^{k_p})}}\right)
    \end{aligned}
\end{equation}
Here, \textit{erfc(.)} is the complementary error function. The effective penalty factor is estimated based on the approximated interactions with the waypoints of all peers of robot-$r$:
\vspace{-0.2cm}
\begin{equation}
\begin{aligned}
\Gamma_r(\mathbf{x}, \mathbf{\hat{X}}_{-r}^{k_r}) &= \prod_{p=1 \wedge p\neq r}^{m}\gamma(\mathbf{x},\mathbf{\hat{x}}_p^{k_p})
\end{aligned}
\end{equation}
\subsection{Information Sharing}
Global inter-robot communication is assumed in this work. However, given the bandwidth limitations of ad-hoc wireless communication (likely in emergency response applications) and its energy footprint of~\cite{li2008robot}, the communication burden needs to be kept low. Thus, along with asynchronous planning, robots share only a down-sampled set of observations. Moreover, each robot broadcasts the following information only after every planning instance: its next planned waypoint ($\mathbf{x}_r^{k_r}$) and observations made over its last path ($\mathcal{D}_r^{k_r}$). 
\section{CASE STUDIES}\label{sec:expts}
\subsection{Distributed Implementation of Bayes-Swarm}
In order to represent the decentralized manner in which {\decBayes} operates, we develop a simulated environment using MATLAB's (R2017b) parallel computing tools, and deployed this environment in a dual 20 core workstation (Intel\textsuperscript{\tiny\textregistered} Xeon Gold 6148 27.5M Cache 2.40 GHz, 20 cores processor, 196 GB RAM). Each robot executes its behavior, as depicted in Fig.~\ref{fig:decBayesflowchart}), in parallel with respect to the rest of the swarm -- updating its own knowledge model after each waypoint and deciding its next waypoint based on its own information and that received from its peers till that point. 

The simulation time step is set at 1 ms. 
The observation frequency over a path is set at 1 Hz. In order to have tractable GP updating, each robot uses all observations dataset ($\mathcal{D}_r^{1:k_r}$) if the size is less than 1,000, otherwise it is down-sampled to 1,000 samples using a simple integer factor. 
%
\begin{figure}[!tp]
\centering
\subfigure[Case study 1: small arena, non-convex bimodal signal distribution] {\includegraphics[width=0.23\textwidth]{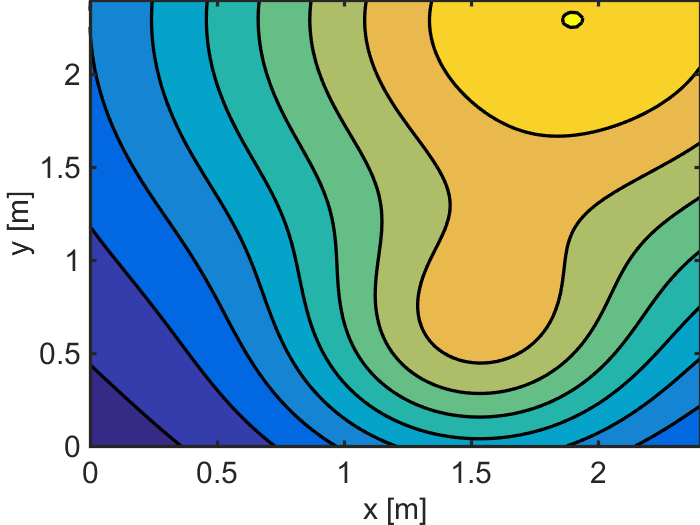}\label{fig:caseStudy2}}\hspace{0.2cm}%
\subfigure[Case study 2: large arena, multi-modal signal distribution]{\includegraphics[width=0.23\textwidth]{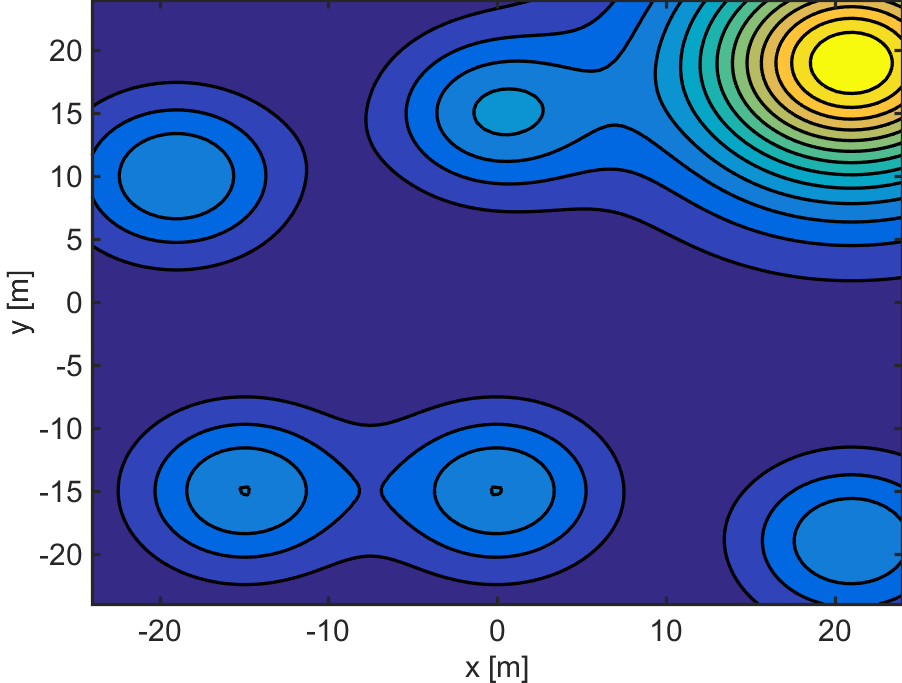}\label{fig:caseStudy4}}\vspace{-0.2cm}
\caption{Two environment cases with different signal distributions.}
\label{fig:caseStudies}
\vspace{-0.6cm}
\end{figure}

\subsection{Case Studies}
To evaluate the \decBayes~algorithm, two types of experiments are conducted using two distinct signal environments. The two environments, shown in Fig.~\ref{fig:caseStudies}, respectively provide a bimodal spatial distribution over a small arena, and a complex multimodal spatial distribution over a larger arena. In {\it{Experiment~1:}} \decBayes~ is run with 5 robots (small swarm size chosen for ease of illustration), to analyse its performance over the two environments, and compare with that of two variations of \decBayes~ (\textit{\decBayes-Sync} and \textit{\decBayes-Explorative}) and an exhaustive search baseline. The synchronized planning (\textit{\decBayes-Sync}) version is implemented by changing the inequality constraint Eq.~\eqref{eq:constraint} to an equality constraint (fixed interval between waypoints) -- to investigate the hypothesized benefits of asynchronous planning. The purely explorative version (\textit{\decBayes-Explorative}) is implemented by using $\alpha = 0$ in Eq. \eqref{eq:mainObjectiveFunc} -- to highlight the need for balancing exploration/exploitation. 
In {\it{Experiment~2:}} a scalability analysis is undertaken to explore the performance of \decBayes~in Case 2, across multiple swarm sizes. 
%
\begin{figure*}[!t]
\centering
\subfigure[$\sigma_1(\mathbf{x})$ at $t$ = 5s (5 samples)]{\includegraphics[width=0.23\textwidth]{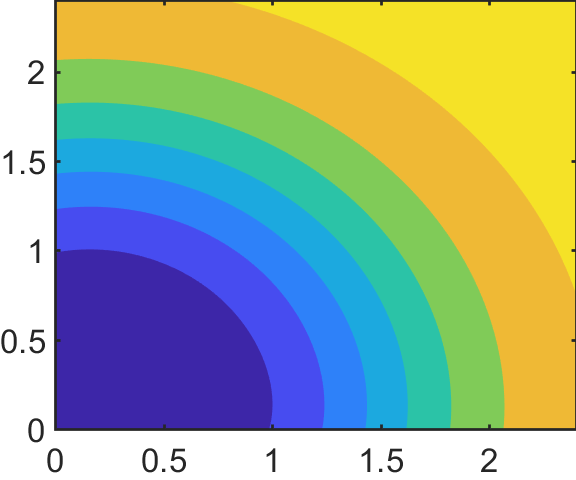}\label{fig:snapshotA}}\hspace{.1cm}%
\subfigure[$\sigma_5(\mathbf{x})$ at $t$ = 5s (25 samples)]{\includegraphics[width=0.23\textwidth]{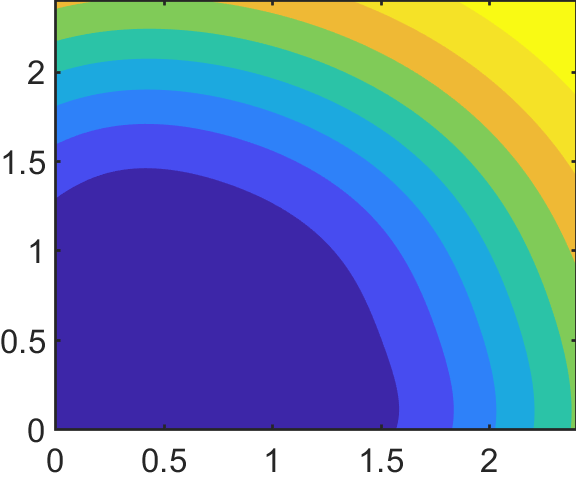}\label{fig:snapshotB}}\hspace{.1cm}%
\subfigure[$\sigma_5(\mathbf{x})$ at $t$ = 20s (100 samples)]{\includegraphics[width=0.23\textwidth]{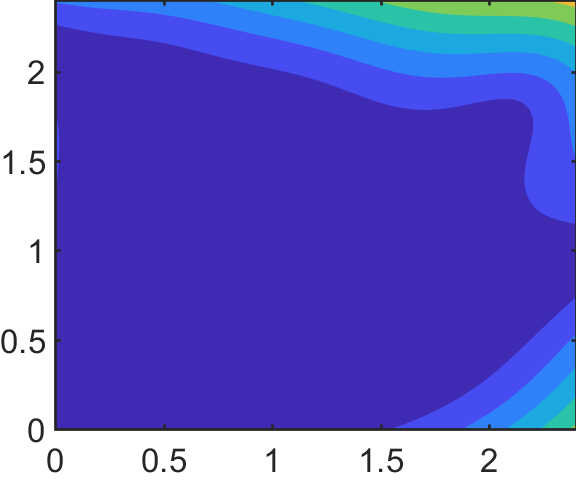}\label{fig:snapshotC}}\hspace{.1cm}%
\subfigure[$\sigma_5(\mathbf{x})$ at $t$ = 36s (180 samples)]{\includegraphics[width=0.23\textwidth]{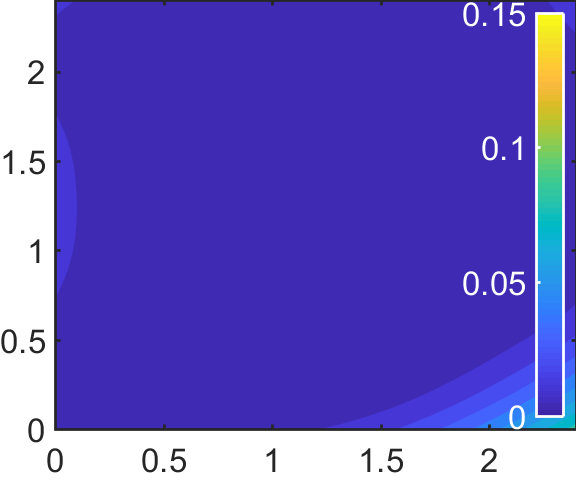}\label{fig:snapshotD}}
\vspace{-0.22cm}
\subfigure[$\mu_1(\mathbf{x})$ at $t$ = 5s]{\includegraphics[width=0.225\textwidth]{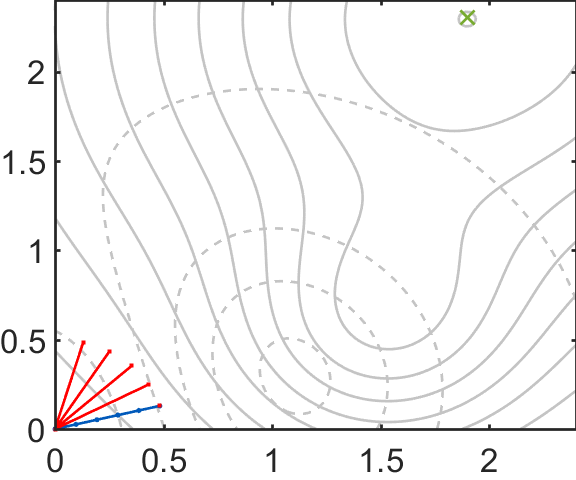}\label{fig:snapshotE}}\hspace{.2cm}%
\subfigure[$\mu_5(\mathbf{x})$ at $t$ = 5s]{\includegraphics[width=0.225\textwidth]{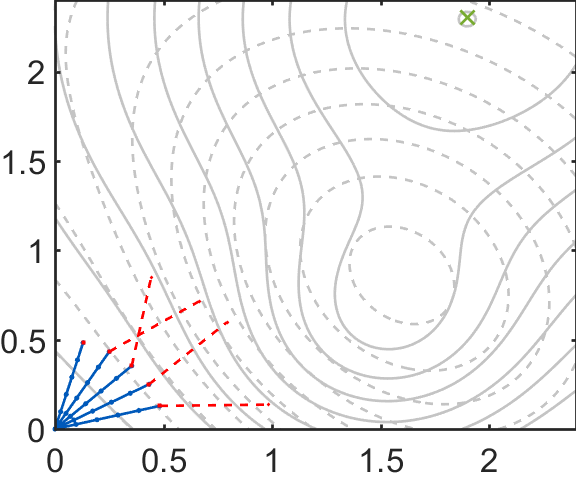}\label{fig:snapshotF}}\hspace{.2cm}%
\subfigure[$\mu_5(\mathbf{x})$ at $t$ = 20s]{\includegraphics[width=0.225\textwidth]{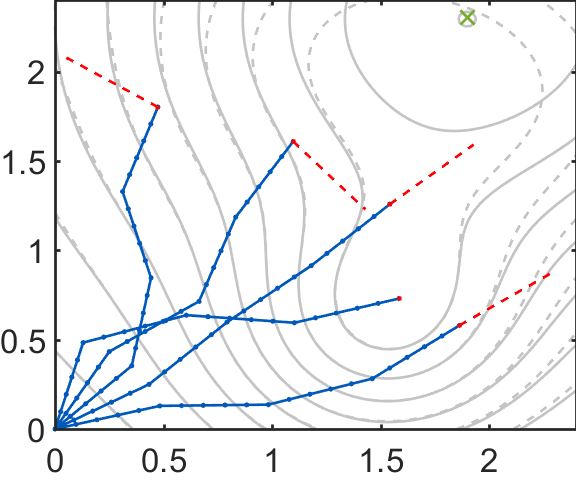}\label{fig:snapshotG}}\hspace{.2cm}%
\subfigure[$\mu_5(\mathbf{x})$ at $t$ = 36s]{\includegraphics[width=0.225\textwidth]{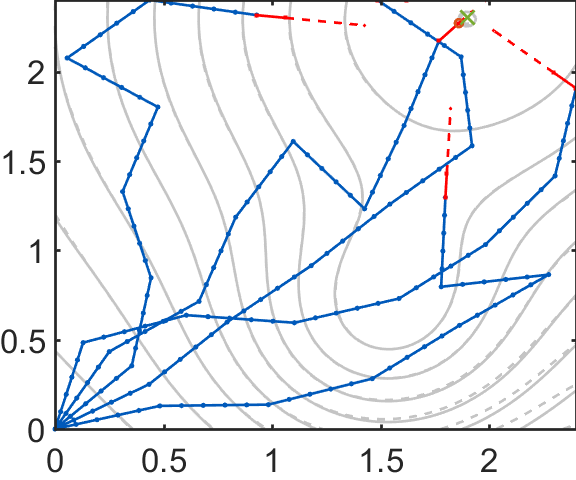}\label{fig:snapshotH}}
\vspace{-0.1cm}
\caption{Experiment 1, case 1: knowledge state and robot path snapshots. \ref{fig:snapshotA}--\ref{fig:snapshotB}: Top figures show the knowledge uncertainty map, in terms of $\sigma_r(\mathbf{x})$). \ref{fig:snapshotE}--\ref{fig:snapshotH}: Bottom figures show robot paths. Blue lines depict travelled paths, observations over which have been shared. Red solid lines depict travelled paths, observations over which have not been shared. Red dashed lines depict planned paths, not yet travelled. Gray solid contours represent the actual signal distribution and gray dashed contours represent the current signal distribution model of the stated robot. The green cross shows the source location.
}
\label{fig:snapshot}
\vspace{-0.7cm}
\end{figure*}

The results of the experiments are evaluated and compared in terms of \textit{relative completion time} ($\tau$) and \textit{mapping error} ($\delta$). The relative completion time represents the search completion time ($t_\text{achieved}$) relative to the idealized completion time ($t_\text{idealized}$), as given by: 
\vspace{-0.2cm}
\begin{equation}
\label{eq:completionTime}
    \tau = \left({t_\text{achieved} - t_\text{idealized}}\right)/{t_\text{idealized}}
\vspace{-0.2cm}
\end{equation}
Here, ${t_\text{idealized}}$ represents the time that a swarm robot would hypothetically take to directly traverse the straight-line path connecting the starting point and the signal source location. Although the focus of this work is source localization (a search problem), Bayes-Swarm can also be applied for mapping purposes (a coverage problem) by setting $\alpha=0$. Hence we report the mapping error $\delta$ (in terms of Root-Mean Square Error  or RMSE), which measures how the response estimated using GP deviates from the actual signal distribution over the arena. The RMSE is computed over a set of $10,000$ test points uniformly distributed over the arena. 

For simulation termination purposes, two criteria are used. The first criterion terminates the search if any robot arrives within $\epsilon$-vicinity of the signal source location. The second criterion terminates the simulated mission, if a maximum allowed search time ($T_\text{max}$) is reached. 
\begin{table}[!b]
\vspace{-0.5cm}
\centering
\caption{Algorithm performance for 5-robot swarm on the two cases}\label{tbl:compareMethods}
\vspace{-.2cm}
\footnotesize
\begin{tabular}{clrr}
\toprule[0.12em]
\textbf{Case} & \textbf{Algorithm} & \textbf{Completion Time} & \textbf{Mapping Error}\\%
\midrule[0.12em]
\multirow{4}{*}{1} & Bayes-Swarm & 0.22 & 0.009\\ 
& Bayes-Swarm-Sync & 0.16 & 0.008\\ 
& Bayes-Explorative & 0.50 & 0.007\\ 
& Exhaustive Search & 6.55 & -\\ 
\midrule[0.12em]
\multirow{4}{*}{2} & Bayes-Swarm & 0.41 & 0.102\\ 
& Bayes-Swarm-Sync & 0.71 & 0.066\\ 
& Bayes-Explorative & 6.11 & 0.054\\ 
& Exhaustive Search &$^\dagger$31.36 & -\\ 
\bottomrule[0.12em]
\end{tabular}
\begin{flushleft}
The maximum allowed search time, the idealized time, robot velocity, and Lipschitz constant are set for both cases as follows: Case 1: $T_\text{max} = 100$s, $T_\text{idealized} = 29.8$s, $T=5$s, $V=0.1$m/s, $M=1$, $L=20$, $\epsilon = 0.05$m; Case 2: $T_\text{max} = 1000$s, $T_\text{idealized} = 140.6$s, $V=0.2$m/s, $T=20$s, $M=1$, $L=200$, $\epsilon = 0.2$m. $^\dagger$ Exhaustive search is done in parallel by 5 robots (2 in 1 qtr, other 3 separately in 3 qtrs of the arena).
\end{flushleft}
\end{table}
%
\section{RESULTS AND DISCUSSION}\label{sec:results}
\subsection{Overall Performance of Bayes-Swarm}\label{ssec:resultsOverall}
Figure~\ref{fig:snapshot} shows snapshots of the status of a 5-robot swarm at different time points, based on the implementation of \decBayes~in the Case 1 environment (Fig. \ref{fig:caseStudy2}). They illustrate how the knowledge uncertainty (top plots), represented by $\sigma_r$ of any given robot-$r$, changes as the swarm robots explore the arena while exchanging information with each other and updating their GP models (bottom plots).

Looking from the perspective of robot 5 -- it can be observed from Figs.~\ref{fig:snapshotF} to ~\ref{fig:snapshotH} how the actions of the swarm helps improve the model of the environment, and correspondingly how the uncertainty in robot 5's knowledge of the signal environment reduces (Figs.~\ref{fig:snapshotB} to ~\ref{fig:snapshotD}). In this Case, the five robots are able to build a relatively accurate model of the environment and find the signal source in 36s (using a total of 180 downsampled measurements); at that point, the estimated and actual signal distributions mostly coincide (Figs.~\ref{fig:snapshotH}). 
The helpful role played by the adaptive weight parameter ($\alpha$) is also evident from Figs. \ref{fig:snapshotG} to \ref{fig:snapshotH}, which show a more explorative behavior early on (e.g., $t=20$s), and a more exploitive behavior later on (e.g., at $t=36$s) when two of the robots converge on the source.

\subsection{Experiment 1: Comparative Analysis of Bayes-Swarm}\label{ssec:resultsComparative}
Table~\ref{tbl:compareMethods} summarizes the performance of the complete, synchronized, and explorative versions of {\decBayes} and the baseline algorithm, in terms of completion time ($\tau$) and mapping error ($\delta$). The results show that \decBayes~and its variations outperform the baseline exhaustive search (by an order of magnitude better efficiency) in both case studies.

Although \textit{Bayes-Swarm} requires greater completion time than \textit{Bayes-Swarm-Sync} in Case 1, the former is clearly superior compared in Case 2. This performance benefit in a complex environment is attributed to the planning flexibility afforded by the absence of synchronization, which enables planning paths of different length across the swarm. 

Our asynchronous implementation, with communication occurring only at waypoints and in a sequence among robots, however introduces non-homogeneity in the models of the environment across the swarm. An example of this is seen from the discrepancies in the knowledge states of robot 1 (Fig. \ref{fig:snapshotA}) and robot 5 (Fig. \ref{fig:snapshotB}) around the 5s timepoint. In the future, this issue could be addressed, while retaining the asynchronous benefits, by designing the communication schedule to be independent of the planning processes.

While the purely explorative version (\textit{Bayes-Explorative}) expectedly provides lower mapping error by reducing the knowledge uncertainty faster, it falls significantly behind both \textit{Bayes-Swarm-Sync} and \textit{Bayes-Swarm} in terms of search completion time, for both environment cases (as evident from Table \ref{tbl:compareMethods}). This illustrates the importance of preserving the exploitation/exploration balance. 

To study the impact of the penalty factor, $\Gamma$, that promotes waypoints away from those planned by peers, we ran {\decBayes} without the penalty. Compared to {\decBayes}, the ``\textit{without penalty factor}" version got stuck in the local signal mode in case 1 and took 1.5 times the time to find the global source in Case 2, with the latter's mapping error performance being also poorer. These observations highlighted the value of the penalty factor.
\vspace{-0.1cm}
\subsection{Experiment 2: Scalability Analysis of Bayes-Swarm}\label{ssec:resultsScalabAnalysis}
\begin{figure}[!t]
\centering
\includegraphics[width=0.45\textwidth]{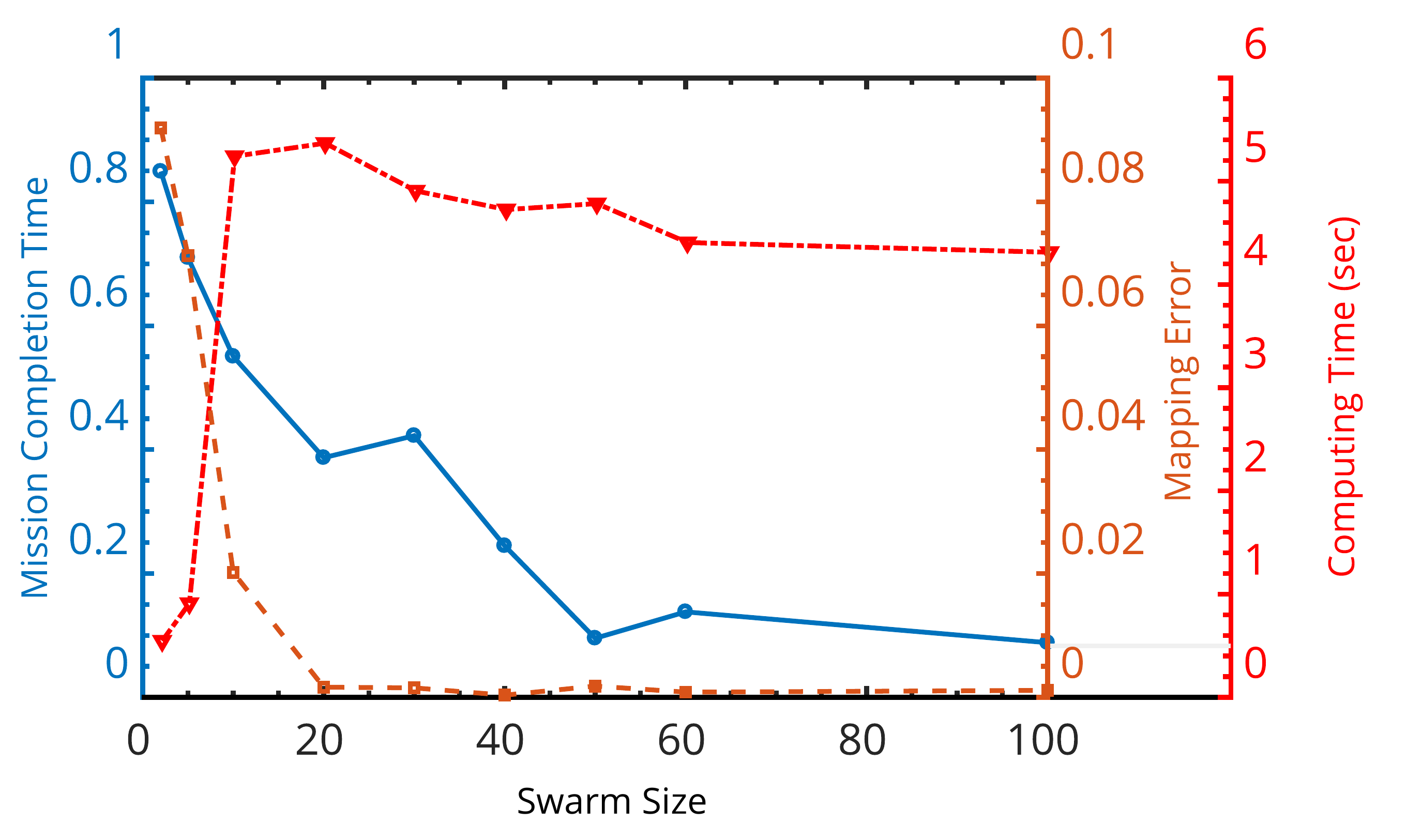}\vspace{-0.4cm}
\caption{Scalability analysis of \decBayes: Variation in performance metrics ($\tau$, $\rho$, computing time) with swarm sizes changing from 2 to 100.}
\label{fig:studying_nrobot}
\vspace{-0.6cm}
\end{figure}

\vspace{-0.1cm}
Here, we run \decBayes~simulations on Case 2 with swarm sizes varying from 2 to 100. Figure~\ref{fig:studying_nrobot} illustrates the results of this study in terms of the relative completion time, mapping error, and computing time per planning instance. The mapping error drops quickly with increasing swarm size, given the resulting increased exploratory capability. The mission completion time also reduces at a remarkable rate between 2 to 50 robots, and then saturates (due to a diminishing marginal utility). Some oscillations are observed in this performance metric, since the penalty characteristics become more aggressive as the swarm size (and thus robot crowding) increases. The computation cost increases, as expected, due to the increasing size of sample sets over which the GP model has to be updated by the robots at each planning instance. This cost is however bounded, via downsampling to a maximum of 1000 samples. Interestingly, the downsampling does not noticeably affect the mission performance improvements. 


%
%

\section{CONCLUSION}
\vspace{-0.1cm}
In this paper, we proposed an asynchronous and decentralized swarm robotic algorithm to perform searching for the maximum strength source of spatially distributed signals. 
To this end, we exploit the batch Bayesian search concept, by making important new modifications necessary to account for the constraints and capabilities that differentiate embodied search from a Bayesian Optimization process. Primarily, a new acquisition function is designed to incorporate the following: 1) knowledge gain over trajectories, as opposed at points; 2) mitigating interactions among planned samples of different robots; 3) time adaptive balance between exploration /exploitation; and 4) accounting for motion constraints.

For evaluation, we used two cases with different arena size and signal distribution complexity. \decBayes~outperformed an exhaustive search baseline by completing the missions 17 times and 76 times faster than exhaustive search, respectively in the simple and the complex cases. The benefits of allowing asynchronous planning and exploitation/exploration balance was also evident in the complex case (noticeably lower mission completion time), studied by setting control experiments (where synchronization and pure exploration was enforced). 

Scalability analysis of \decBayes~demonstrated a somewhat superlinear reduction in completion time and mapping error with increasing number of robots. The computing cost per waypoint planning did increase sharply with increasing swarm size, since swarm size exacerbates the cost of re-fitting the GP (onboard swarmbots), which grows exponentially along the mission as more samples are collected. Efficient refitting approximations, e.g., particle learning, will be explored in the future to address this concern. Another direction of future research would be consideration of partial observation (attributed to communication constraints), which along with physical demonstration would allow more comprehensive appreciation of the {\decBayes} algorithm.

\appendices
\section{Gaussian Process Model}\label{sec:GP}
Gaussian process (GP) models provide non-parametric surrogates~\cite{rasmussen2003gaussian} that can be used for Bayesian inference over a function space ~\cite{snoek2012practical}. For a set of $n$ observations, $\mathcal{D}={\mathbf{x}_i,y_i|i=1\dots n}$, GP expresses the observed values $y_i$ as a summation of the approximating function $f(\mathbf{x}_i)$ and an additive noise $\epsilon_i$, i.e., $y_i = f(\mathbf{x}_i) + \epsilon_i$. 
Assuming the noise follows an independent, identically distributed Gaussian distribution with zero mean and variance, $\sigma^2_n$, we have $\epsilon \sim \mathcal{N}(0,\sigma_n^2)$.
The function $f(\mathbf{x})$ can then be estimated by a GP with mean $\mu(\bx)$ and a covariance kernel $\sigma^2(\bx)$: 
\vspace{-0.2cm}
\begin{equation}\label{eq:GP}
    f \; | \; \mathbf{x}, \; \mathbf{X}, \; \mathbf{y} \sim \mathcal{N}\big(\mu(\mathbf{x},\mathbf{X},\mathbf{y}),\sigma^2(\mathbf{x},\mathbf{X})\big)
    \vspace{-0.2cm}
\end{equation}
\begin{align}
\mu(\mathbf{x},\mathbf{X},\mathbf{y}) = \mathbf{k}_n(\mathbf{x})^T [\mathbf{K} +\sigma_n^2(\mathbf{x})\mathbf{I}]^{-1}(\mathbf{y}-\Phi\beta)\\
\sigma^2(\mathbf{x},\mathbf{X}) = k(\mathbf{x},\mathbf{x}) - \mathbf{k}_n(\mathbf{x})^T [\mathbf{K} +\sigma_n^2(\mathbf{x})\mathbf{I}]^{-1}\mathbf{k}_n(\mathbf{x})
\end{align}
Here $\Phi$ is the vector of explicit basis functions and $\mathbf{K} = \mathbf{K}(\mathbf{X},\mathbf{X}|\theta)$ is the covariance matrix, $(\mathbf{K})_{ij} = k(\mathbf{x}_i, \mathbf{x}_j)$, with $\mathbf{k}_n(\mathbf{x}) = [k(\mathbf{x}_1,\mathbf{x}), \dots, k(\mathbf{x}_n,\mathbf{x})]^T$. In this paper, the squared exponential kernel is used to define the covariance $k(\mathbf{x}_i,\mathbf{x})$.
The GP hyper-parameters are determined by maximizing the log-likelihood $P$ as a function of $\beta,\theta,\sigma_n^2$, i.e.,
\vspace{-0.2cm}
\begin{equation}
\hat{\beta},\,\hat{\theta},\,\hat{\sigma}_n^2 = \text{arg}\max_{\beta,\theta,\sigma_n^2}\log{P(\mathbf{y}|\mathbf{X},\beta,\theta,\sigma_n^2)}
\end{equation}
\vspace{-0.2cm}where\vspace{-0.1cm}
\begin{equation}
\begin{aligned}
\log{P(\mathbf{y}|\mathbf{X},\beta,\theta,\sigma_n^2)} = &-\frac{1}{2}(\mathbf{y}-\Phi\beta)^T\Lambda(\mathbf{x})^{-1}(\mathbf{y}-\Phi\beta)\\
&- \frac{N_s}{2}\log{2\pi}-\frac{1}{2}\log{|\Lambda(\mathbf{x})|}
\end{aligned}
\end{equation}

\bibliographystyle{IEEEtran}
\bibliography{payam2018mas}

\end{document}